\documentclass[twocolumn,prd]{revtex4}

\usepackage{amsmath,amsthm,amsfonts,amscd,amssymb} 
\usepackage{eucal}
\usepackage{braket}
\usepackage{slashed}
\usepackage{hyperref}
\usepackage{graphicx}
\usepackage{bm}

\usepackage{amsmath,amsthm,amsfonts,amscd,amssymb} 
\usepackage{eucal}
\usepackage{braket}
\usepackage{slashed}
\usepackage{hyperref}
\usepackage{graphicx}
\usepackage{bm}

\newcommand{\Hi}{\mathcal{H}}
\newcommand{\Oi}{\mathcal{O}}

\newcommand{\be}{\begin{equation}}
\newcommand{\ee}{\end{equation}} 
\newcommand{\mb}{\mathbf}

\newcommand{\bear}{\begin{eqnarray}}
\newcommand{\eear}{\end{eqnarray}}

\DeclareMathOperator{\tr}{tr}

\begin{document}

\title{Dressed infrared quantum information}

  \author{Daniel Carney}
  \email{carney@umd.edu}
  \altaffiliation[Current address: ]
{Joint Center for Quantum Information and Computer Science, University of Maryland, College Park, MD and the Joint Quantum Institute, National Institute of Standards and Technology, College Park, MD.}
       \affiliation{Department of Physics and Astronomy, University of British Columbia, BC, Canada}
  \author{Laurent Chaurette}
  \email{dodeca@phas.ubc.ca}
       \affiliation{Department of Physics and Astronomy, University of British Columbia, BC, Canada}
  \author{Dominik Neuenfeld}
  \email{dneuenfe@phas.ubc.ca}
       \affiliation{Department of Physics and Astronomy, University of British Columbia, BC, Canada}
  \author{Gordon Walter Semenoff}
  \email{gordonws@phas.ubc.ca}
       \affiliation{Department of Physics and Astronomy, University of British Columbia, BC, Canada}

\begin{abstract}
We study information-theoretic aspects of the infrared sector of quantum electrodynamics, using the dressed-state approach pioneered by Chung, Kibble, Faddeev-Kulish and others. In this formalism QED has an IR-finite $S$-matrix describing the scattering of electrons dressed by coherent states of photons. We show that measurements sensitive only to the outgoing electronic degrees of freedom will experience decoherence in the electron momentum basis due to unobservable photons in the dressing. We make some comments on possible refinements of the dressed-state formalism, and how these considerations relate to the black hole information paradox.
\end{abstract}


\maketitle

There are two common methods for dealing with infrared divergences in quantum electrodynamics. One is to form inclusive transition probabilities, tracing over arbitrary low-energy photon emission states.\cite{Bloch:1937pw,suura,Weinberg:1965nx} However, one may wish to retain an $S$-matrix description instead of working directly with probabilities. To this end, a long literature initiated by Chung, Kibble, and Faddeev-Kulish has advanced a program in QED where one forms an infrared-finite $S$-matrix between states of charges ``dressed'' by long-wavelength photon modes.\cite{Dirac:1955uv,dollard,Chung:1965zza,kibble,Kulish:1970ut,Zwanziger:1974jz,Kapec:2017tkm} The extension to perturbative gravity in flat spacetime has been initiated in \cite{Ware:2013zja}.

In the inclusive probability formalism, one is forced to trace out soft photons to get finite answers. In previous work, we showed that this leads to an almost completely decohered density matrix for the outgoing state after a scattering event.\cite{Carney:2017jut} This paper analyses the situation in dressed state formalisms, in which no trace over IR photons is needed to obtain a finite outgoing state. However, consider the measurement of an observable sensitive only to electronic and high-energy photonic degrees of freedom. We show that for such observables, there will be a loss of coherence identical to that obtained in the inclusive probability method. Quantum information is lost to the low-energy bremsstrahlung photons created in the scattering process.

The primary goal of this paper is to give concrete calculations exhibiting the dressed formalism and how it leads to decoherence. To this end, we work with the formulas from the papers of Chung and Faddeev-Kulish. The result of this calculation should carry over identically to any of the existing refinements of Chung's formalism. In section \ref{discussion}, we make a number of remarks on possible refinements to the basic dressing formalism, give an expanded physical interpretation of our results, and relate our work to literature in mathematical physics on QED superselection rules. In section \ref{blackholes} we make remarks on how this work fits into the recent literature on the black hole information paradox; in brief, we believe that our results are consistent with the recent proposal of Strominger \cite{Strominger:2017aeh}, but not the original proposal of Hawking, Perry and Strominger.\cite{Hawking:2016msc,Hawking:2016sgy}

\section{IR-safe $S$-matrix formalism}
\label{formalism}

Following Chung, we study an electron with incoming momentum $\mb{p}$ scattering off a weak external potential. This $1 \to 1$ process is simple and sufficient to understand the basic point; at the end of the next section, we show how to generalize our results to $n$-particle scattering. The electron spin will be unimportant for us and we supress it in what follows. The standard free-field Fock state $\ket{\mb{p}}$ for the electron is promoted to a dressed state $\ket{\tilde{\mb{p}}}$ as follows. For a given photon momentum $\mb{k}$ we define the soft factor
\be
\label{softfactor}
F_{\ell}(\mb{k},\mb{p}) = \frac{p \cdot e_{\ell}(\mb{k})}{p \cdot k} \phi(\mb{k},\mb{p}).
\ee
Here $\ell = 1,2$ labels the photon polarization states, and $\phi(\mb{k},\mb{p})$ is any function that smoothly goes to $\phi \to 1$ as $|\mb{k}| \to 0$. We introduce an IR regulator (``photon mass'') $\lambda$ and an upper infrared cutoff $E > \lambda$, which can be thought of as the energy resolution of a single-photon detector in our experiment. Let
\be
R_{\mb{p}} = e \sum_{\ell=1}^{2} \int_{\lambda < |\mb{k}| < E} \frac{d^3 \mb{k}}{\sqrt{2k}} \left[ F_{\ell}(\mb{k},\mb{p}) a_{\ell}^{\dagger}(\mb{k}) - F_{\ell}^{*}(\mb{k},\mb{p}) a_{\ell}(\mb{k}) \right] 
\ee
and define the single-electron dressing operator
\begin{align}
\begin{split}
\label{dressingop}
W_{\mb{p}} & = \exp \left\{ R_{\mb{p}} \right\} \\
& = N_{\mb{p}} \exp \left\{ e \sum_{\ell = 1}^{2} \int \frac{d^3\mb{k}}{\sqrt{2k}} F_{\ell}(\mb{p},\mb{k}) a_{\ell}^{\dagger}(\mb{k}) \right\} \\
& \times \exp \left\{ -e \sum_{\ell = 1}^{2} \int \frac{d^3\mb{k}}{\sqrt{2k}} F^*_{\ell}(\mb{p},\mb{k}) a_{\ell}(\mb{k}) \right\},
\end{split}
\end{align}
where in the second line, we have put this coherent-state displacement operator into its normal-ordered form, with normalization factor \footnote{This factor diverges, so these states have zero norm. In this sense, the dressed-state formalism simply re-organizes the calculations such that the divergences are in the definitions of the states $\ket{\tilde{\mb{p}}}$ instead of the $S$-matrix elements. We view this is a major difficulty with these formalisms, and understanding this better would be very useful. See eg. \cite{Dybalski:2017mip} for some ideas in this direction.}
\be
N_{\mb{p}} = \exp \left\{ -\frac{e^2}{2} \sum_{\ell = 1}^{2} \int \frac{d^3\mb{k}}{2k} \left| F_{\ell}(\mb{p},\mb{k}) \right|^2 \right\}.
\ee
Here and in the following all momentum-space integrals are evaluated in the shell $\lambda < |\mb{k}| < E$. The dressed single-electron state $\ket{\tilde{\mb{p}}}$ is then defined as
\be
\ket{\tilde{\mb{p}}} = W_{\mb{p}} \ket{\mb{p}}.
\ee
This consists of the electron and a coherent state of on-shell, transversely-polarized photons. 

Consider now an incoming dressed electron scattering into a superposition of outgoing dressed electron states. The outgoing state is, to lowest order in perturbation theory in the electric charge,
\begin{align}
\begin{split}
\label{outstate}
\ket{\psi} = \int d^3\mb{q} \tilde{S}_{\mb{q} \mb{p}} \ket{\tilde{\mb{q}}}.
\end{split}
\end{align}
At higher orders there will be additional photons in the outgoing state; as explained in the next section, these will not affect the infrared behavior studied here, so we ignore them for now. Here the $S$-matrix is just the standard Feynman-Dyson time evolution operator, evaluated between dressed states. That is,
\be
\tilde{S}_{\mb{q} \mb{p}} = \braket{ \tilde{\mb{q}} | S | \tilde{\mb{p}}},
\ee
with $S = T \exp \left( -i \int_{-\infty}^{\infty} V(t) dt \right)$ as usual.\cite{Weinberg:1995mt} As calculated by Chung, the dressed $1 \to 1$ elements of this matrix are independent of the IR regulator $\lambda$ and thus infrared-finite as we send $\lambda \to 0$. We can write the matrix element
\be
\tilde{S}_{\mb{q} \mb{p}} = \left( \frac{E}{\Lambda} \right)^{A} S_{\mb{q} \mb{p}}^{\Lambda}
\ee
where
\be
A = - \frac{e^2}{8\pi^2} \beta^{-1} \ln \left[ \frac{1+\beta}{1-\beta} \right], \ \ \ \beta = \sqrt{1 - \frac{m^4}{(p \cdot q)^2}}.
\ee
The undressed $S$-matrix element on the right side means the amplitude computed by Feynman diagrams with photon loops evaluated only with photon energies above $\Lambda$ and evaluated between undressed electron states, that is, with no external soft photons. By definition, this quantity is infrared-finite; moreover, the dependence on the scale $\Lambda$ cancels between the prefactor and $S^{\Lambda}$.

\section{Soft radiation and decoherence}

The state \eqref{outstate} is a coherent superposition of states, each containing a bare electron and its corresponding photonic dressing. The presence of hard photons in the outgoing state will not change our conclusions below, so for simplicity we ignore them. In particular, the density matrix formed from this state has off-diagonal elements of the form
\be
\tilde{S}^*_{\mb{q}' \mb{p}} \tilde{S}_{\mb{q} \mb{p}} \ket{\tilde{\mb{q}}} \bra{\tilde{\mb{q}}'} .
\ee
These states have highly non-trivial photon content. However, if one is doing a measurement involving only the electron degree of freedom, then these photons are unobserved, and we can make predictions with the reduced density matrix of the electron, obtained by tracing the photons out. The resulting electron density matrix has coefficients damped by a factor involving the overlap of the photon states, namely
\be
\label{electronDM}
\rho_{electron} = \int d^3\mb{q} d^3\mb{q}' \tilde{S}^*_{\mb{q}' \mb{p}} \tilde{S}_{\mb{q} \mb{p}} D_{\mb{q} \mb{q}'} \ket{\mb{q}} \bra{\mb{q}'}
\ee
where the dampening factor is given by the photon-vacuum expectation value
\be
D_{\mb{q} \mb{q}'} = \braket{ 0 | W_{\mb{q}'}^{\dagger} W_{\mb{q}} | 0 }.
\ee
Straightforward computation gives this factor as
\begin{align}
\begin{split}
\label{damping}
D_{\mb{q} \mb{q}'} & = \exp \left\{ -\frac{e^2}{2} \sum_{\ell=1}^{2} \int \frac{d^3\mb{k}}{2k} \left| F_{\ell}(\mb{q}) - F^*_{\ell}(\mb{q}') \right|^2 \right\} \\
& = \exp \left\{ -e^2 \int \frac{d^3\mb{k}}{2k} \frac{(q-q')^2}{(q \cdot k) (q' \cdot k)} \right\}.
\end{split}
\end{align}
In this integrand, since $q$ and $q'$ are two timelike vectors with the same temporal component, we have that the numerator is positive definite and the denominator is positive. It is therefore manifest that we have $D = 1$ if $q = q'$ and $D = 0$ otherwise, since the integral over $d^3\mb{k}$ diverges in its lower limit. Thus, tracing the photons leads to an electron density matrix that is completely diagonalized in momentum space. 

It is noteworthy that the factor \eqref{damping} depends only on properties of the outgoing superposition; it has no dependence on the initial state. This may seem surprising since we are tracing over outgoing radiation, the production of which depends on both the initial and final electron state. The point is that the damping factor measures the distinguishability of the radiation fields given the processes $\mb{p} \to \mb{q}$ and $\mb{p} \to \mb{q}'$. The radiation field for a scattering process consists of two pieces added together: a term $A_{\mu} \sim p_{\mu}/p \cdot k$ peaked in the direction of the incoming electron and a term $A_{\mu} \sim q_{\mu}/q \cdot k$ peaked in the direction of the outgoing electron. The outgoing radiation fields with outgoing electrons $q,q'$ are then only distinguishable by the second terms here, since both radiation fields will have the same pole in the incoming direction.

The damping factor \eqref{damping} is precisely what was found in \cite{Carney:2017jut}, reduced to the problem of $1 \to 1$ scattering. The mechanism is the same: physical, low-energy photon bremsstrahlung is emitted in the scattering. These photons are highly correlated with the electron state and thus, if one does not observe them jointly with the electron, one will measure a highly-decohered electron density matrix. The only difference is bookkeeping: in the dressed formalism, the bremsstrahlung photons are folded into the dressed electron states (the incoming/outgoing parts of the bremsstrahlung in the incoming/outgoing dressing, respectively). However, referring to ``an electron'' as a state including these soft photons is just an abuse of semantics. In an actual measurement of the electron momentum, one does not measure these soft photons.

The dressed states are not energy eigenstates, and in fact contain states of arbitrarily high total energy. This should be contrasted with the inclusive-probability treatment used by Weinberg, which has a cutoff on both the single-photon energy $E$ and the total outgoing energy contained by all the photons $E_T \geq E$ in the outgoing state.\cite{Weinberg:1965nx} This additional parameter, however, appears only in the ratio $E_T/E$ in Weinberg's probability formulas, and one finds that the dependence on $E_T$ vanishes as $E_T \to \infty$. This can be understood because what is important is the very low-energy behavior of the photons, so moving an upper cutoff has limited impact.

We note that \eqref{outstate} does not include effects from the bremsstrahlung of additional soft photons beyond those in the dressing. There is no kinematic reason to exclude such photons, so the outgoing state should properly be written as
\begin{align}
\begin{split}
\label{outstate2}
\ket{\psi} = \sum_{n=0}^{\infty} \sum_{\{\ell\}} \int d^3\mb{q} d^{3n}\mb{k} \tilde{S}_{\mb{q} \{ \mb{k} \ell \};\mb{p}} \ket{\tilde{\mb{q}}}.
\end{split}
\end{align}
Here $\{\mb{k} \ell\} = \{ \mb{k}_1 \ell_1, \ldots, \mb{k}_n \ell_n\}$ is a list of $n$ photon momenta and polarizations. By the dressed version of the soft photon factorization theorem (see appendix), we have that
\be
\tilde{S}_{\mb{q} \mb{k} \ell; \mb{p}} =  \tilde{S}_{\mb{q} \mb{p}} \times e \Oi \left( |\mb{k}|^0 \right),
\ee
or in other words $\lim_{|\mb{k}| \to 0} |\mb{k}| \tilde{S}_{\mb{q} \mb{k} \ell; \mb{p}} = 0$. Thus, when we take a trace over $n$-photon dressed states in \eqref{outstate2}, we obtain a sum of additional decoherence factors of the form
\begin{align}
\begin{split}
D^{nm}_{\mb{q} \mb{q}'} & = e^{n+m} \Oi\left(|\mb{k}|^0\right) \times \sum_{\ell_1, \ldots, \ell_n} \sum_{\ell'_1, \ldots, \ell'_m} \int d^{3n}\mb{k} d^{3m} \mb{k}' \\
& \braket{ 0 | a_{\ell'_m}(\mb{k}'_m) \cdots a_{\ell'_1}(\mb{k}'_1) W_{\mb{q}'}^{\dagger} W_{\mb{q}} a^{\dagger}_{\ell_1}(\mb{k}_1) \cdots a^{\dagger}_{\ell_n}(\mb{k}_n)  | 0 }.
\end{split}
\end{align}
Evaluating the inner product using \eqref{dressingop}, one finds
\be
D^{nm}_{\mb{q} \mb{q}'} \sim \left[ \sum_{\ell=1}^{2} \int \frac{d^3{\mb{k}}}{\sqrt{2 k}} \text{Re} \left( F_{\ell}(\mb{q}) - F_{\ell}(\mb{q}') \right) \right]^{n+m},
\ee
which is infrared-finite. Summing these contributions, which exponentiate, will not change the conclusion that \eqref{damping} leads to vanishing off-diagonal electron density matrix elements.

Finally, we explain the generalization to $n$-electron states. We will find that the same decoherence is found in the dressed formalism as in the inclusive formalism.\cite{Carney:2017jut} Following Faddeev-Kulish \cite{Kulish:1970ut}, we write the multi-particle dressing operator by replacing \eqref{dressingop} with
\begin{align}
\begin{split}
R_p & \to e \sum_{l=1}^2 \int \frac{d^3 \mathbf p}{(2\pi)^3} \int \frac{d^3 \mathbf k}{\sqrt{2k}} \\
& \left[ F_l(\mathbf{k},\mathbf{p}) a_l^\dagger(\mathbf k) - F^*_l(\mathbf{k},\mathbf{p}) a_l^\dagger(\mathbf k) \right] \rho(\mathbf p),
\end{split}
\end{align} 
where we have introduced an operator which counts charged particles with momentum $\mathbf p$.
\begin{align}
\rho(\mathbf p) = \sum_s \left( b_{\mathbf p,s}^\dagger b_{\mathbf p,s} - d_{\mathbf p,s}^\dagger d_{\mathbf p,s}\right),
\end{align}
and the $b$ and $d$ are electron and positron operators, respectively.\footnote{Note that in the multi-particle case there is an infinite phase factor which needs to be included in the definition of the S-matrix. Since this phase factor does not affect our discussion, we ignore it in the following.} As in the one-particle case, additional photons do not affect the IR behaviour of scattering amplitudes. Hence, we will ignore them and only consider the case where the out-state is a linear superposition of dressed electron states. In that case we have to replace the outgoing momentum by list of momenta, $\mb{q} \to \beta = \{ \mb{q}_1, \mb{q}_2, \ldots \}$ and similarly $\mb{q}' \to \beta' = \{ \mb{q}'_1, \mb{q}'_2, \ldots \}$. This results to a replacement in \eqref{damping} of 
\begin{align}
\begin{split}
F_{\ell}(\mb{q}) & \to \sum_{n \in \beta} F_{\ell}(\mb{q}_n) \\
F^*_{\ell}(\mb{q}') & \to \sum_{m \in \beta'} F^*_{\ell}(\mb{q}'_m).
\end{split}
\end{align}
Using the explicit form of $F$ in the limit $\mb{k} \to 0$, the damping factor \eqref{damping} then then becomes
\begin{align}
\label{fulldamping}
D_{\beta\beta'} = \exp \left[ -e^2 \int \frac{d^3 \mathbf k}{2k} \sum_{m,n \in \beta,\beta'} \frac{\eta_m \eta_n p_m \cdot q_n}{(q_m \cdot k)(q_n \cdot k)} \right].
\end{align}
In this equation the labels $m,n$ both run over the full set $\beta \cup \beta'$, and  $\eta_n = +1$ if $n \in \beta$ while $\eta_n = -1$ if $n \in \beta'$. This is precisely the quantity $\Delta A_{\beta\beta', \alpha}$ defined in \cite{Carney:2017jut}, so we see that the results of that paper carry over to the dressed formalisms used here.

\section{Physical interpretation}
\label{discussion}

Dressed-state formalisms are engineered to provide infrared-finite transition amplitudes, as opposed to inclusive probabilities constructed in the traditional approach studied in \cite{Carney:2017jut}. In the dressed formalism, the outgoing state \eqref{outstate} is a coherent superposition of states $\ket{\tilde{\mb{p}}}$ consisting of electrons plus dressing photons. However, if one does a measurement of an observable sensitive only to the electron state, the measurement will exhibit decoherence because the unobserved dressing photons are highly correlated to the electron state. We have given a concrete calculation showing that the damping factor \eqref{fulldamping} is identical in either the dressed or undressed formalism.

The physical relevance of this calculation rests on the idea that the basic observable is a simple electron operator in Fock space. What would be much better would be to use a dressed LSZ reduction formula to understand the asymptotic limits of electron correlation functions. \cite{Zwanziger:1974jz,Dybalski:2017mip} Nevertheless, the basic physical picture seems clear: in a scattering experiment, one does not measure an electron plus a finely-tuned shockwave of outgoing bremsstrahlung photons, just the electron on its own. This is responsible for well-measured phenomena like radiation damping.

QED has a complicated asymptotic Hilbert space structure which is still somewhat poorly understood.  For example, although Faddeev-Kulish try to define a single, separable Hilbert space $\Hi_{as}$ \cite{Kulish:1970ut,Dybalski:2017mip} other authors have argued that one needs an uncountable set of separable Hilbert spaces.\cite{kibble,Zwanziger:1974jz} Formally, this is related to the fact that the dressing operator \eqref{dressingop} does not converge on the usual Fock space. A related idea is that one can argue that QED has an infinite set of superselection rules based on the asymptotic charges
\be
\label{qedcharges}
Q(\Omega) = \lim_{r \to \infty} r^2 E_r(r,\Omega)
\ee
defined by the radial electric field at infinity.\cite{Gervais:1980bz,Buchholz:1982ea} We believe that the calculations presented here and in \cite{Carney:2017jut} demonstrate the physical mechanism for enforcing such a superselection rule. The charges \eqref{qedcharges}, the currents defined in our previous work \cite{Carney:2017jut}, and the large-$U(1)$ charges defined in \cite{Kapec:2015ena,Campiglia:2015qka} are presumably closely related, and working out the precise relations is an interesting line of inquiry.

\section{Black hole information}
\label{blackholes}

The recent resurgence of interest in infrared issues in QED and gravity was sparked largely by a proposal of Hawking, Perry and Strominger suggesting that information apparently ``lost'' in the process of black hole formation and evolution could be encoded in soft radiation.\cite{Hawking:2016msc,Hawking:2016sgy} The original proposal was that there are symmetries which relate ``hard'' scattering (like the black hole process) to soft scattering and thus led to constraints on the $S$-matrix. As emphasized by a number of authors, this is not true in the dressed state approach.\cite{Mirbabayi:2016axw,Gabai:2016kuf,Gomez:2016hxz,Bousso:2017dny} As we review in the appendix, soft modes decouple from the dressed hard scattering event at lowest order, in the sense that $\lim_{\omega \to 0} [a_{\omega},S_{dressed}] = 0$. Dropping a soft boson into the black hole will not yield any information about the black hole formation and evaporation process.

However, a more recent proposal due to Strominger is to simply posit that outgoing soft radiation purifies the outgoing Hawking radiation.\cite{Strominger:2017aeh} That is, the state after the black hole has evaporated is of the form $\ket{\psi} = \sum_{a} \ket{a}_{Hawking} \ket{a}_{soft}$, such that the Hawking radiation is described by a thermal density matrix, i.e. $\rho_{Hawking} = \tr_{soft} \ket{\psi} \bra{\psi} \approx \rho_{thermal}$. We believe that both the results presented here and those in our previous work are consistent with this proposal. In either the inclusive or dressed formalism, the final state of any scattering process contains soft radiation which is highly correlated with the hard particles  because the radiation is created due to accelerations in the hard process. The open issue is to explain why the hard density matrix coefficients behave thermally, which likely relies on details of the black hole $S$-matrix.

\section{Conclusions}

When charged particles scatter, they experience acceleration, causing them to radiate low-energy photons. If one waits an infinitely long time (as mandated by an $S$-matrix description), these photons cause severe decoherence of the charged particle momentum state. This was first seen in \cite{Carney:2017jut} in the standard formulation of QED involving IR-finite inclusive cross section, and here we have shown the same conclusion holds in IR-safe, dressed formalisms of QED; they should carry over in a simple way to perturbative quantum gravity. These results constitute a sharp and robust connection between the infrared catastrophe and quantum information theory, and should provide guidance in problems related to the infrared structure of gauge theories.

\section*{Acknowledgements}

We thank Colby Delisle, Wojciech Dybalski, Philip Stamp, and Jordon Wilson for discussions. All of us are grateful for support from NSERC. DC was additionally supported by the Templeton Foundation and the Pacific Institute of Theoretical Physics, and DN by a UBC Four Year Doctoral Fellowship and the Simons Foundation.

\begin{figure*}[!ht]
\includegraphics[scale=.7]{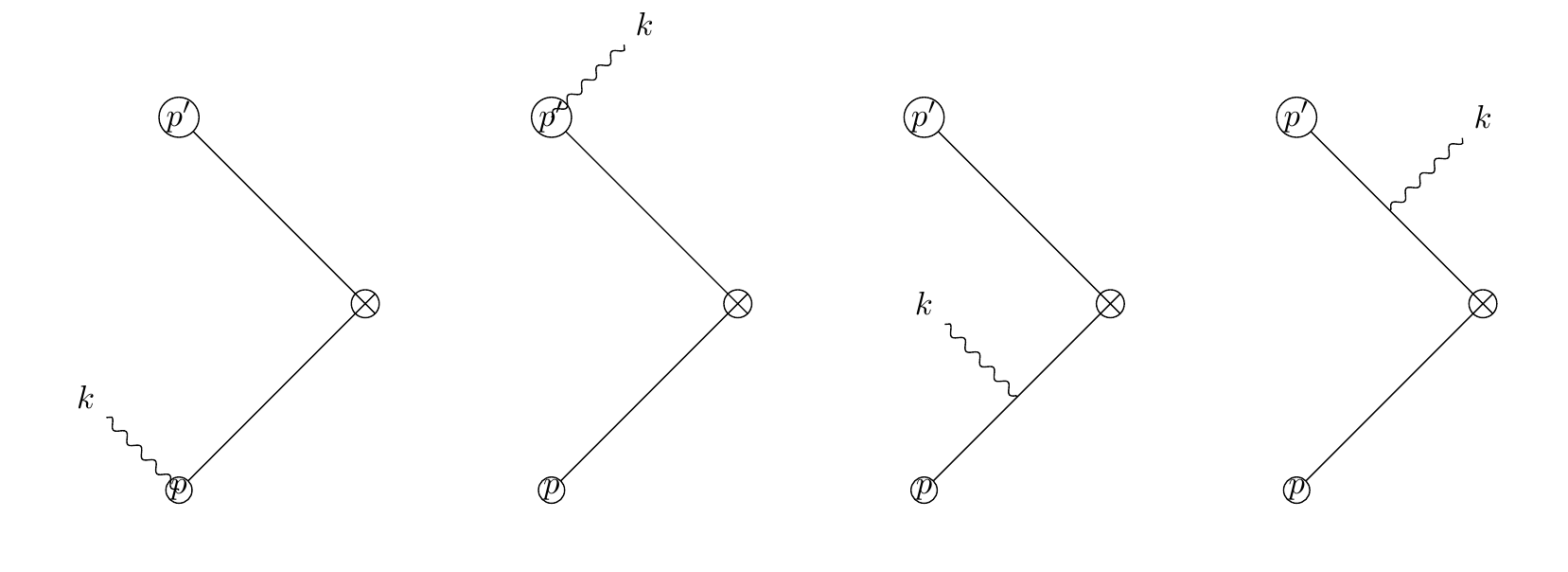}
\caption{Diagrams contributing to the dressed scattering with additional bremsstrahlung. The first two diagrams correspond to mixing of the emitted photon and dressing photons (circles), while the latter two diagrams correspond to the usual Feynman diagrams where the photon is emitted from the electron lines.}
\label{diagrams}
\end{figure*}

\appendix
\label{factorizationapp}
\section{Dressed soft factorization} The soft photon theorem looks somewhat different in dressed QED. In standard, undressed QED, the theorem says that the amplitude for a process $\mb{p} \to \mb{q}$ accompanied by emission of an additional soft photon of momentum $\mb{k}$ and polarization $\ell$ has amplitude 
\be
\label{undressed1photon}
S_{\mb{q} \mb{k} \ell, \mb{p}} = e \left[ \frac{q \cdot e^*_{\ell}(\mb{k})}{q \cdot k} - \frac{p \cdot e^*_{\ell}(\mb{k})}{p \cdot k} \right]  S_{\mb{q}, \mb{p}}.
\ee
This is singular in the $k \to 0$ limit. On the other hand, in the dressed formalism of QED, the statement is that
\be
\label{dressed1photonappendix}
\tilde{S}_{\mb{q} \mb{k} \ell, \mb{p}} = e f(\mb{k}) \tilde{S}_{\mb{q}, \mb{p}},
\ee
where $f(\mb{k}) \sim \Oi(|\mb{k}|^0)$, so that the right-hand side is finite as $k \to 0$. We can see this by straightforward computation. In computing the matrix element \eqref{dressed1photonappendix}, there will be four Feynman diagrams at lowest order in the charge (see fig. \ref{diagrams}). We will get the usual pair of Feynman diagrams coming from contractions of the interaction Hamiltonian with the external photon state, leading to the poles \eqref{undressed1photon}. Moreover we will get a pair of terms coming from contractions of the interaction Hamiltonian with dressing operators. These contribute a factor
\begin{align}
\begin{split}
\label{dressedadditions}
& \left[ F^*_{\ell}(\mb{k},\mb{p}) - F^*_{\ell}(\mb{k},\mb{q}) \right] \\ 
& \to \left[ \frac{q \cdot e^*_{\ell}(\mb{k})}{q \cdot k} - \frac{p \cdot e^*_{\ell}(\mb{k})}{p \cdot k} \right] + \Oi(|\mb{k}|^0),
\end{split}
\end{align}
times $-e$, where the limit as $k \to 0$ follows from the definition \eqref{softfactor}. This extra contribution precisely cancels the poles in \eqref{undressed1photon}, leaving only the order $\Oi(|\mb{k}|^0)$ term.

\bibliography{dressed-iqi-references}

\end{document}